\documentclass[a4paper,aps,prd,onecolumn,preprintnumbers,showpacs,nofootinbib]{revtex4}
\usepackage{amsmath,amssymb,graphics,epsfig,subfigure}
\usepackage{color}
\usepackage{graphicx}

\newcommand{\be}{\begin{equation}}
\newcommand{\bea}{\begin{eqnarray}}
\newcommand{\eea}{\end{eqnarray}}
\newcommand{\ba}{\begin{array}}
\newcommand{\ea}{\end{array}}
\newcommand{\ee}{\end{equation}}

\def\bse{\begin{subequations}}
\def\ese{\end{subequations}}

\begin{document}

\renewcommand{\baselinestretch}{1.3}
\title{Boson star with  parity-odd symmetry  in wormhole spacetime}

\author{Yuan Yue$^{1}$}
\author{Peng-Bo Ding$^{2}$}
\author{Yong-Qiang Wang$^{2}$}\email{yqwang@lzu.edu.cn}
\affiliation{$^{1}$College of Mathematics and Computer Science, Northwest Minzu University, Lanzhou, 730030, China\\
$^{2}$School of Physical Science and Technology, Lanzhou University, Lanzhou 730000, China}

\begin{abstract}
In this paper, we revisit  the model of   bosonic matter in the form of a free complex scalar field
with a nontrivial wormhole spacetime topology supported by
a free phantom field.   We obtain a new type of boson star with wormhole solutions,   in which the complex scalar field possess full parity-odd symmetry with respect to the two asymptotically flat
spacetime regions. When the size of the throat  is   small,
The behavior of boson stars with wormhole approaches that of boson stars.  When the size of the throat is intermediate,  the typical spiraling dependence
of the mass and the particle number on the frequency of the boson stars is replaced by a loop structure. However, as the size becomes relatively large,   the loop structure will   also disappear.  In particular,  The complex scalar field   could form two boson stars  with opposite phase differences with respect to the two
spacetime regions  in the limit of vanishing throat
size. We analyze the properties of this new type of boson stars with wormhole and  further show
that the wormhole  spacetime geometry.

\end{abstract}

% \Keywords{ Large Extra dimensions, braneworlds, localization}

%\pacs{04.50.-h, 11.27.+d }

% 04.50.+h Gravity in more than four dimensions, Kaluza-Klein theory,
           % unified field theories, alternative theories of gravity
           %(see also 11.25.M Compactification and four-dimensional models), dilaton gravity
% 11.27.+d Extended classical solutions; cosmic strings,
           %domain walls, texture (see 98.80.C in cosmology)

\maketitle

\section{Introduction}
Wormholes are fascinating hypothetical structures in space that have captured the imagination of scientists and science-fiction writers alike. The concept of a wormhole was first introduced in 1916 by the German physicist Ludwig Flamm. However, it wasn't until 1935 that A. Einstein and N. Rosen developed the theory of wormholes as part of their work on general relativity, also known as an Einstein-Rosen bridge \cite{Einstein:1935tc}, which is a theoretical tunnel or shortcut that connects two separate points in spacetime. These tunnels can be used to travel vast distances through space or even time, bypassing the need to traverse the vast distances of the universe at sub-light speeds. However,  such a Einstein-Rosen  wormhole would not be traversable for matter or for light \cite{Kruskal:1959vx,Fuller:1962zza}. Later,
M. Morris and K. Thorne \cite{Morris:1988cz} found that creating or finding a   humanly traversable wormhole would require the existence of exotic matter with negative energy density, which violates null energy conditions (NEC) \cite{Visser:1989kh}.
Several types of such traversable wormholes have actually been discovered in \cite{Ellis:1973yv,Ellis:1979bh,Bronnikov:1973fh,Kodama:1978dw}, and  a phantom scalar field is selected  as the potential candidate for such exotic matter, which has a reversed sign in its kinetic term.

Boson stars (BSs) originated from the study of geons (gravitational electromagnetic entity), which were proposed by Wheeler~\cite{Wheeler:1955zz,Power:1957zz}. Geons are considered to be stable, particle-like solutions without horizons in the context of gravity coupled to a classical electromagnetic field and  gravitational wave. However, geons consisting of massless vector fields have not been found. Later, Kaup~{\em et al.} obtained Klein-Gordon geons (i.e., boson stars) by replacing the massless vector field with a massive complex scalar field~\cite{Kaup:1968zz}. Ruffini also independently studied boson stars by considering quantized real scalar fields~\cite{PhysRev.187.1767}. The original boson stars are spherically symmetric and composed of free scalar fields with fundamental configurations.
It is generalized to the cases of rotation~\cite{Schunck:1996,Schunck:1996he,1997PhRvD..56..762Y}, the excited BSs~\cite{Bernal:2009zy,Collodel:2017biu,Wang:2018xhw}, static multipolar BSs~\cite{Herdeiro:2020kvf,Herdeiro:2021mol}, the construction of vector boson stars~(Proca stars) - see~{\em e.g.}~\cite{Brito:2015pxa,Herdeiro:2019mbz,Minamitsuji:2018kof} and different kinds of multi-field or multi-state configuration~\cite{Bernal:2009zy,Alcubierre:2018ahf,Li:2019mlk,Li:2020ffy,Sanchis-Gual:2021edp,Dzhunushaliev:2021vwn,Henriques:1989ez,Zeng:2021oez,Sun:2022duv,Liang:2022mjo,Ma:2023vfa}.
There are other generalizations as well, see the reviews~\cite{Schunck:2003kk,Liebling:2012fv}.

Previously, most of the research on boson stars focused on topologically trivial spacetime.
In \cite{Dzhunushaliev:2014bya,Hoffmann:2017jfs},  the investigation focused on mixed systems that included boson stars with wormholes in the core of the star.
The bosonic matter that comprised these systems consisted of an ordinary complex boson field with self-interaction, which could exhibit symmetric or asymmetrical distribution across the two asymptotically flat regions. As a result, the wormhole spacetime  could be either symmetric or asymmetric.
Furthermore,   the rotation of the scalar field induces a symmetric and asymmetric rotating wormhole spacetime as well \cite{Hoffmann:2017vkf,Hoffmann:2018oml}.
In these  works, both the symmetric and asymmetric wormhole solutions   the complex scalar field exhibit the same sign difference with respect to the two asymptotically flat spacetime regions connected by the wormhole. Thus, it is an interesting question whether there exist complex scalar fields with opposite sign differences at the two ends of a wormhole.
 In this paper,  we reinvestigate the model of   bosonic matter in the form of a free complex scalar field
with a nontrivial wormhole spacetime topology supported by
a free phantom field.  A new kind of  boson star with wormhole solutions  is obtained,  in which the complex scalar field could possess full parity-odd symmetry with respect to the two asymptotically flat
spacetime regions.  In particular,  the scalar field  that form two boson stars with respect to the two
spacetime regions has opposite phase differences in the limit of vanishing throat
size.  It is noteworthy that this new kind of solution still possesses symmetric wormhole spacetime structure.

   The paper is organized as follows. In Sec. \ref{model} , we review the model of  a   complex scalar  field and a  phantom scalar  field
field coupled with gravity.  We present the boundary conditions and quantities of interest in Sec. \ref{sec3},   discuss the numerical results in
Sec. \ref{sec4}. The last section is devoted to conclusion.

\section{The Model}\label{model}

We consider the Einstein-Hilbert action including  the Lagrangian for the   complex scalar  field and the phantom scalar  field
field
\begin{equation}\label{action}
  S=\int\sqrt{-g}d^4x\left(\frac{R}{2\kappa}+\mathcal{L}_{p}+\mathcal{L}_{m}\right),
\end{equation}
where $R$ is the Ricci scalar.  The term
$\mathcal{L}_{p}$ and $\mathcal{L}_{m}$ are  the Lagrangians defined by
with
\begin{eqnarray}
\mathcal{L}_{m}  &= & -\nabla_a\Psi^*\nabla^a\Psi  - \mu_0^2\Psi\Psi^*,  \\
\mathcal{L}_{p}  & =   &   \nabla_a\Phi\nabla^a\Phi \ .
\end{eqnarray}
Here $\Psi$  and $\Phi$ represent the complex scalar  field and the phantom   field, respectively.
By varying the action (\ref{action}) with respect to the metric, we can obtain the Einstein equations
\begin{equation}
  \label{eq:EKG1}
R_{\mu\nu}-\frac{1}{2}g_{\mu\nu}R-\kappa T_{\mu\nu}=0 \ ,
\end{equation}
with stress-energy tensor
\begin{equation}
T_{\mu\nu} = g_{\mu\nu}({{\cal L}}_m+{{\cal L}}_p)
-2 \frac{\partial ({{\cal L}}_m+{{\cal L}}_p)}{\partial g^{\mu\nu}} \ ,
\label{eq1}
\end{equation}
and the matter field equations by varying with respect to the scalar field and  phantom field
\begin{equation}
  \label{eq:EKG2}
  \Box\Psi-\mu_0^2\Psi=0,
\end{equation}
and
\begin{equation}
  \label{eq:EKG2}
  \Box\Phi=0,
\end{equation}

We consider the  general static spherically symmetric solution with a wormhole,
and adopt the Ans\"atzes as follows, see {\em e.g.}~\cite{Hoffmann:2017jfs}:
\begin{equation}  \label{line_element1}
 ds^2 = -e^{A}  dt^2 +B e^{-A}   \left[ d r^2 + h (d \theta^2+\sin^2 \theta d\varphi^2)   \right]\,,
\end{equation}
where $A$ and $B$   are functions of  radial coordinate $r$,  $h=r^2+r_0^2$ with  the throat parameter  $r_0$.
 and $r$  ranges from positive infinity to negative infinity.
 The two limits $r\rightarrow \pm\infty$ correspond to two distinct  asymptotically flat spacetime.
Furthermore, we assume stationary complex scalar  field and phantom field in the form
\begin{eqnarray}  \label{an2}
  \Psi&=\psi(r)e^{i\omega t}, \;\;\;\;  \Phi&=\phi(r).
\end{eqnarray}
Here, $\psi$ is only a real function of the radial coordinate $r$, and the constant $\omega$ is referred to as the synchronization frequency. Moreover, the phantom field $\Phi$ is also a real function and is independent of the time coordinate $t$.

Substituting the above Ans\"atzes  (\ref{line_element1}) and (\ref{an2}) into the Einstein equations (\ref{eq:EKG1}) leads to the following field equations
\begin{eqnarray}
4 \kappa  e^{-2 A}  \psi^2 B^2 \left(\mu_0^2  e^{A}-\omega^2\right)+\frac{3 r B'}{h}+B''-\frac{B'^2}{2 B}=0,  \label{ew1}\\
A''+\frac{2 r A'}{h}+\frac{A' B'}{2 B}+2 \kappa  e^{-2 A}  \psi^2 B \left(\mu_0^2  e^{A}-2 \omega^2\right)=0,   \label{ew2}\\
\frac{e^{2 A} \left(-B^2 \left(h^2 A'^2+4 r_0^2\right)+4 r h B B'+h^2 B'^2\right)}{2 h^2 B^2}
+\kappa  \left(2  \psi^2 B \left(\mu_0^2 e^{A}-\omega^2\right)+e^{2 A} \left(\phi'^2-2 \psi'^2\right)\right)=0. \label{ew3}
\end{eqnarray}
Variation of the action with respect to the
complex scalar field and to the phantom field
leads to the equations
\begin{eqnarray}
\left(\omega^2 e^{-2 A}  \psi B-\mu_0^2 e^{-A} B\right)+\frac{(h \sqrt{B} \psi')'}{\sqrt{B} h}
& = &
0 \ ,
\label{eqSk}\\
(h \sqrt{B} \phi')' & = & 0 .
\label{eqPh}
\end{eqnarray}
These five equations are divided into three groups: three of
these equations (\ref{ew1}), (\ref{ew2}) and (\ref{eqSk}) are solved together, yielding a coupled system of three ODEs on
the three unknown functions $A$, $B$ and $\psi$.    After obtaining the solution for these three functions,  ones can solve Eq. (\ref{eqPh})  to determine the value of $\phi$.  The remaining one  equation (\ref{ew3}) is treated as the constraint and used to
check the numerical accuracy of the method.
%(The $\varphi\varphi$ and $\theta\theta$ equations are equivalent.)
Especially,
The  above derivative in Eq. (\ref{eqPh})  happens to be zero, which means it can be equal to a constant.
Integrating the last equation,   ones can  leads to
\begin{equation}
\phi' = \frac{\sqrt{\cal D}}{h \sqrt{B}}\ .
\label{phip}
\end{equation}
Here the constant of integration $\cal D$ denotes the scalar charge of the phantom field.
By taking Eq.~(\ref{phip}) into Eq.~(\ref{ew3})
the scalar charge $\cal D$
can be expressed into the following form

\begin{eqnarray}
{\cal D}
  = -\frac{h^2 B}{\kappa  e^{2 A}  }\left( \frac{e^{2 A} \left(-B^2 \left(h^2 A'^2+4 r_0^2\right)+4 r h B B'+h^2 B'^2\right)}{2 h^2 B^2}
+\kappa  \left(2  \psi^2 B \left(\mu_0^2 e^{A}-\omega^2\right)-2 e^{2 A}  \psi'^2\right)\right).
\label{eqD2}
\end{eqnarray}

The system of coupled non-linear equations (\ref{ew1}), (\ref{ew2}) and (\ref{eqSk}) presents a formidable challenge in seeking a general analytical solution, necessitating the use of numerical methods to obtain the solutions to differential equations. However, prior to numerically solving the aforementioned equations, it is worthwhile to investigate two special solution cases. Firstly, when the complex scalar field vanishes, one can derive the solution for an Ellis wormhole.  Secondly, as the throat size $r_0$ approaches zero, two mini-boson stars emerge about the $r_0=0$ symmetry and correspond to two distinct asymptotically flat spacetimes. These two special cases provide valuable insight into the behavior of the system and can inform subsequent numerical analyses.

\section{Boundary conditions and quantities of interest }\label{sec3}
%%%%%%%%%%%%%%%%%%%%%%%%%%%%%%%%%%%%%%%%%%%%%%%%%%%%%%%%%%
To solve   the   coupled  non-linear  equations   (\ref{ew1}), (\ref{ew2}) and (\ref{eqSk})   in asymptotically flat spacetime,
we give appropriate boundary conditions for the functions as follows
\begin{eqnarray}
\psi=A=0, \hspace{5pt}   B=1
\end{eqnarray}
at infinity~($r \rightarrow \infty$).
At the origin ($r=0$),
the symmetry of   functions can be used to distinguish between two types of solutions.
For even parity solutions, we require the following boundary conditions
\begin{align}
\partial_r \psi=\partial_r A=\partial_r B= 0 \ .
\end{align}
The above boundary conditions
  were employed in \cite{Hoffmann:2017jfs} to investigate the scenario of symmetric solutions.
Considering that   the complex scalar   with  parity-odd symmetry,  we impose
\begin{align}
\psi= 0 \ , \nonumber \\
\partial_r A=\partial_r B= 0 \ .
\end{align}

The ADM mass $M$   is  the key quantities we are interested in,
which is encoded in the asymptotic expansion of metric components
\begin{eqnarray}
g_{tt}= -1+\frac{2 M}{r}+\cdots \ .
\end{eqnarray}
The action of the complex scalar field is invariant under the $U(1)$ transformation $\psi\rightarrow e^{i\alpha}\psi$ with a constant $\alpha$. According to Noether's theorem, there is a conserved current corresponding to the complex scalar field:
\begin{equation}\label{equ9}
 J^{\mu} = -i\left(\psi^*\partial^\mu\psi - \psi\partial^\mu\psi^*\right), \;\;\;\;\;\;\; J^\mu_{\,\,\,; \mu} =0 \;.
\end{equation}
Integrating the timelike component of the above conserved currents on a spacelike hypersurface $\cal{S}$, ones could obtain Noether charge:
\begin{eqnarray}
Q  &=& \int_{\cal S}J_S^t \nonumber \\
&= &- \int J^t \left| g \right|^{1/2} dr d\Omega_{2}
\nonumber \\
&=& 8 \pi \omega \int_0^{\infty} |g| ^{1/2}   \frac{\psi^2}{e^{A}} \,dr \, \ .
\label{Qc}
\end{eqnarray}
The global charge $Q$ is equivalent to the number of particles in the complex scalar field.

\section{Numerical results}\label{sec4}
\begin{figure}[]
  \begin{center}
  \includegraphics[width=7.8cm]{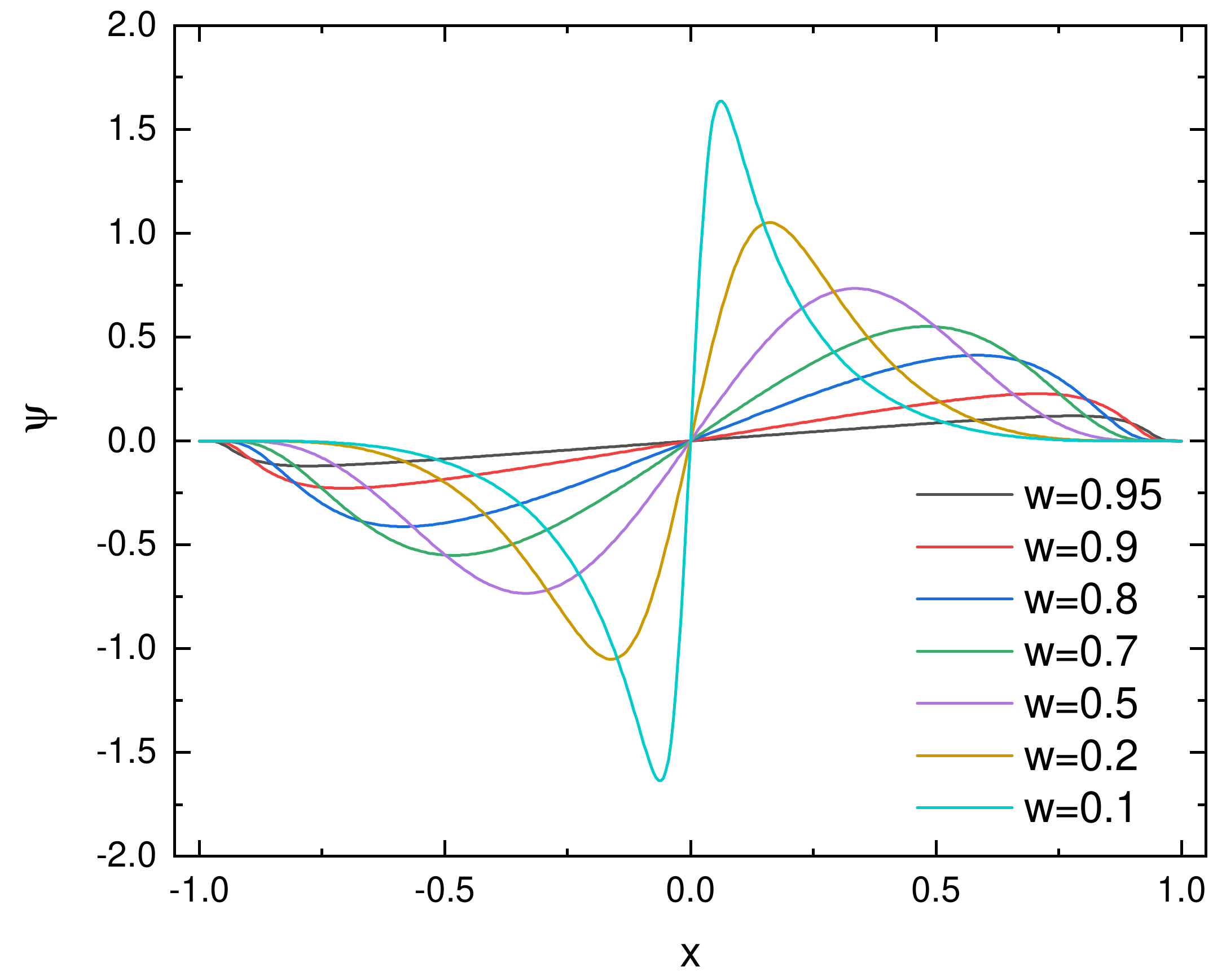}
   \includegraphics[width=7.8cm]{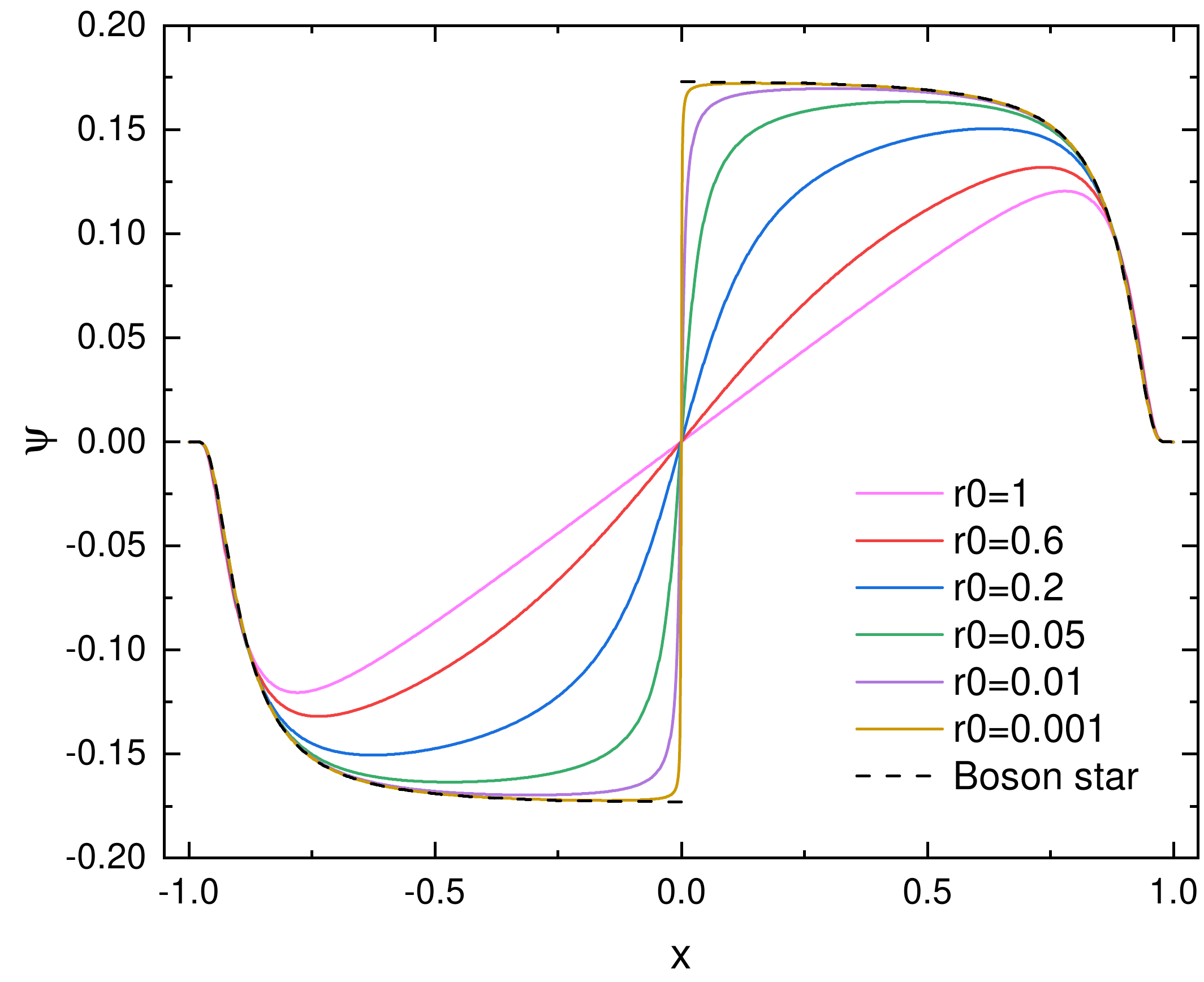}
  \end{center}
  \caption{The radial distribution of the complex scalar field. \textit{Left}: The distribution  with several values of the  frequency $\omega$
for  $r_0=1$. \textit{Right}: The distribution with several values of the  throat size $r_0$
for   $\omega=0.95$.
  }\label{phase}
\end{figure}
In this work, all the numbers are dimensionless as follows
\begin{eqnarray}
r \rightarrow r\mu_0 \hspace{5pt}, \hspace{5pt} \phi \rightarrow \phi \kappa^{-1/2}\hspace{5pt}, \hspace{5pt} \omega_i \rightarrow \omega_i/\mu_0 \hspace{5pt}.
\end{eqnarray}
For convenience,
we set $ \mu_0 = 1$ and $\kappa=0.2$.
We transform the radial coordinates by the following equation
\begin{eqnarray}
\label{transform}
x= \frac{2}{\pi}\arctan(r) \;,
\end{eqnarray}
The use of Eq.~(\ref{transform}) map the infinite region $(-\infty,\infty)$ to the finite region $(-1,1)$.
Next, what we need is to discretize the differential equations on a grid.
This allows the ordinary differential equations to be approximated by algebraic equations.
The grid with $500$ points covers the integration region $-1 \leq x \leq 1$.
In our work, relative errors are less than $10^{-6}$.

First, as an example, we plot in Fig. \ref{phase} two typical
profiles of our numerical results for the scalar field $\psi$ as a
function of  $x$.
In the left panel,  the throat size  $r_0 =1$ is fixed and the curves  with several values of the frequency $\omega$ are shown. We can see that the scalar field $\psi$ has an antisymmetric structure about the origin $x=0$.  $\psi_{max}$ (the maximum value of the field function $\psi$) increases  as the frequency   $\omega$ decreases.  Moreover, the extreme points converge towards the origin $x=0$ as the frequency decreases.
In the right panel,  the frequency  $\omega = 0.95$ is fixed and curves for different values of  throat size $r_0$ are displayed.
In particular, we represent the two boson star solutions with black dashed lines, and these two solutions form an anti-symmetric structure at the two ends of the wormhole.
 Ones can see that at positions where $x$ is large, the curves for different values of $r_0$ do not differ much and basically overlap.  In the region where $x$ is small, the value of the scalar field increases as $r_0$ decreases.
When the limit $ r_0
\rightarrow0$, the solutions that are very close to two boson stars  at the two ends of the wormhole except near the origin.

\subsection{Global charges of the
solutions}
\begin{figure}[]
  \begin{center}
  \includegraphics[width=8.3cm]{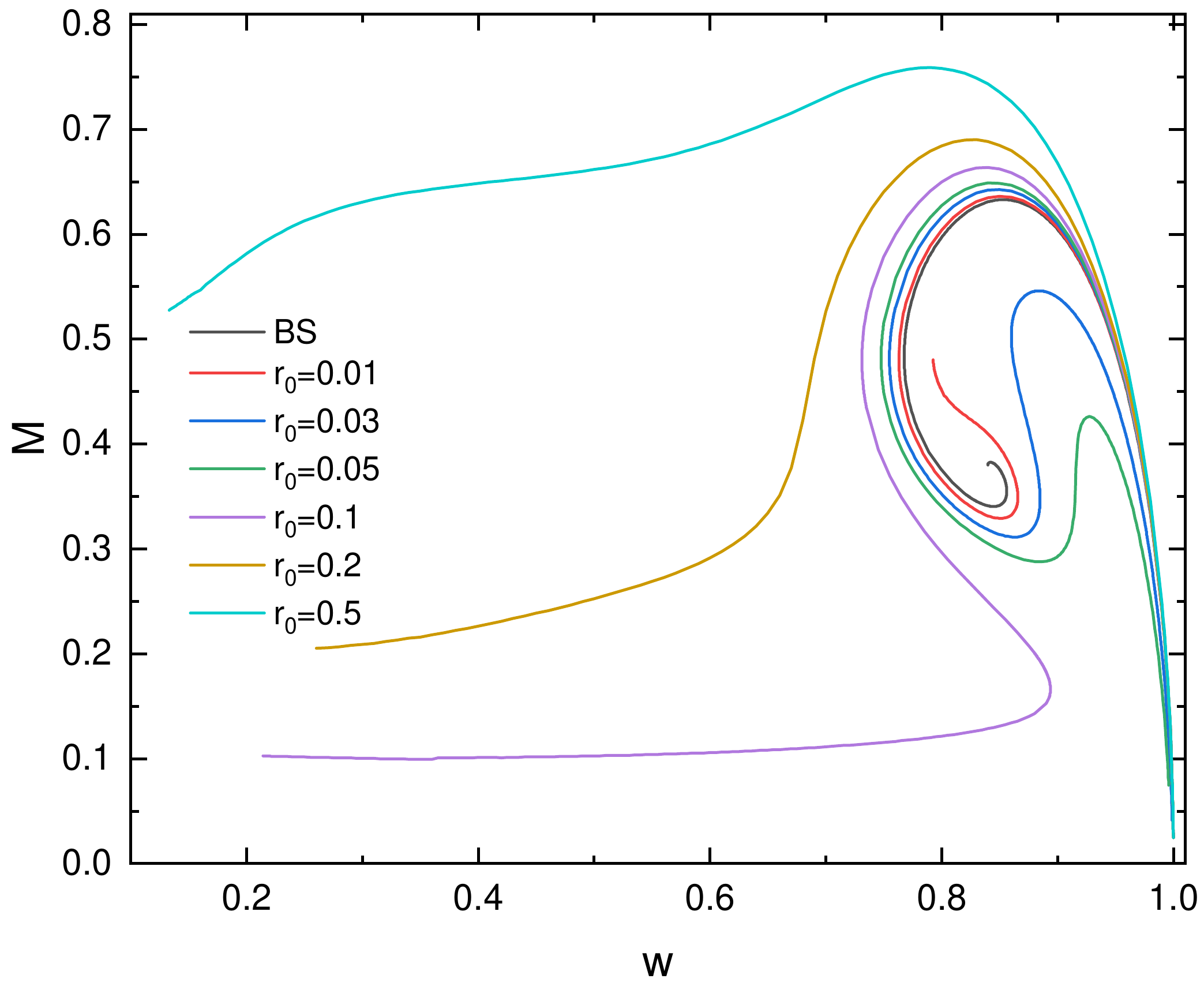}
   \includegraphics[width=8.3cm]{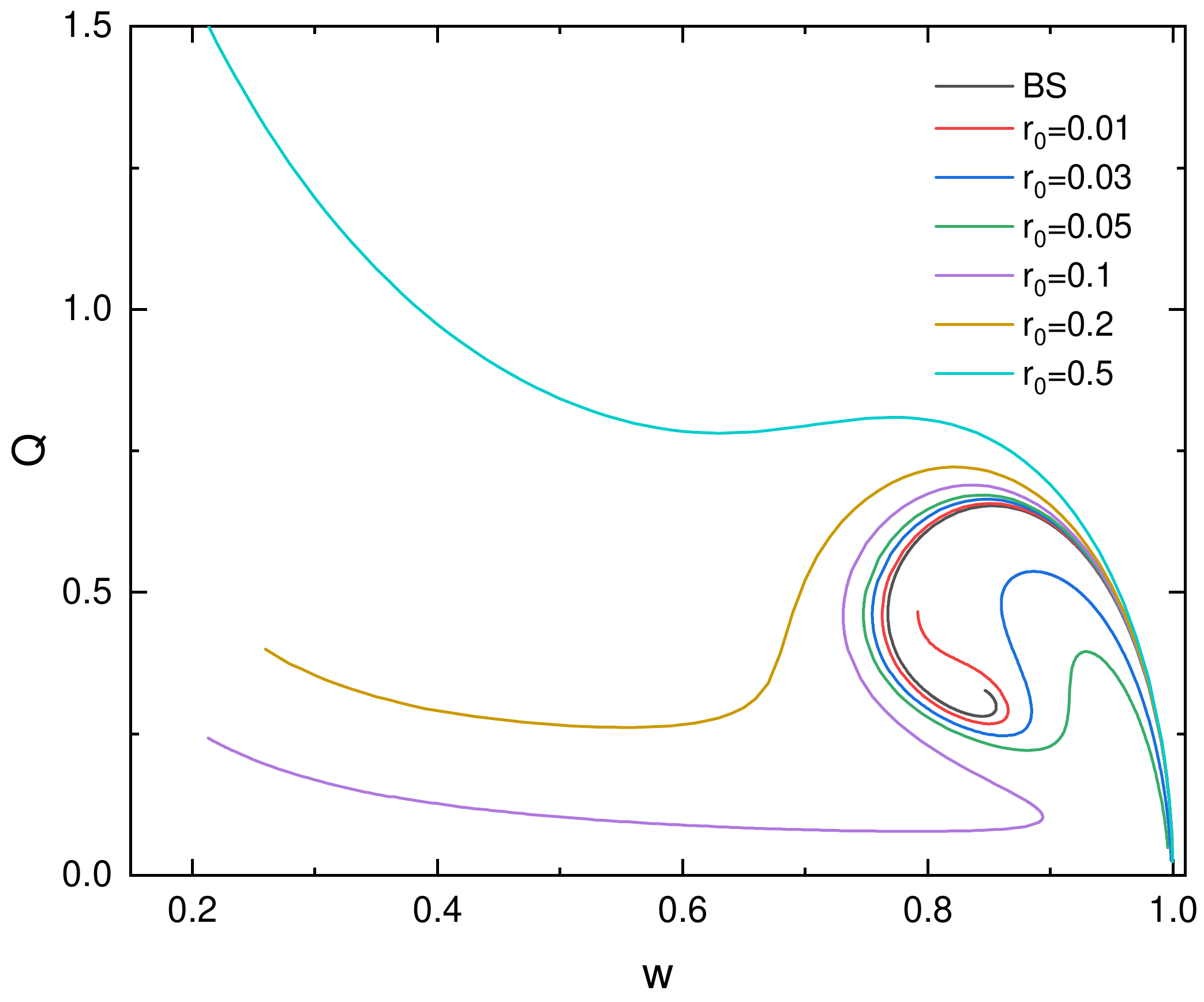}
  \end{center}
  \caption{The ADM mass  $M$ and Noether charge $Q$   as a function of the   frequency $\omega$ with several values of the  throat size $r_0$.
  }\label{phase2}
\end{figure}

In this subsection, we will initially examine the global charges of the resulting wormhole solutions. In Fig. \ref{phase2}, we exhibit the mass $M$  and Noether charges $Q$ versus
the frequency $\omega$ with the different values of  throat size $r_0$,
the black curve corresponds to the boson star solution.
 It is interesting that the curve does
not form a spiral.   Instead, When $r_0$ starts to increase from 0 value,  the curve will first form a loop structure. Then, as $r_0$ continues to increase, the multi-valued curve will become a single-valued curve.
In the first branch, when the frequency is fixed, we find that as $r_0$ increases, both the mass $M$ and charge $Q$ also increase, which have similar behavior.

\begin{figure}[]
  \begin{center}
  \includegraphics[width=11cm]{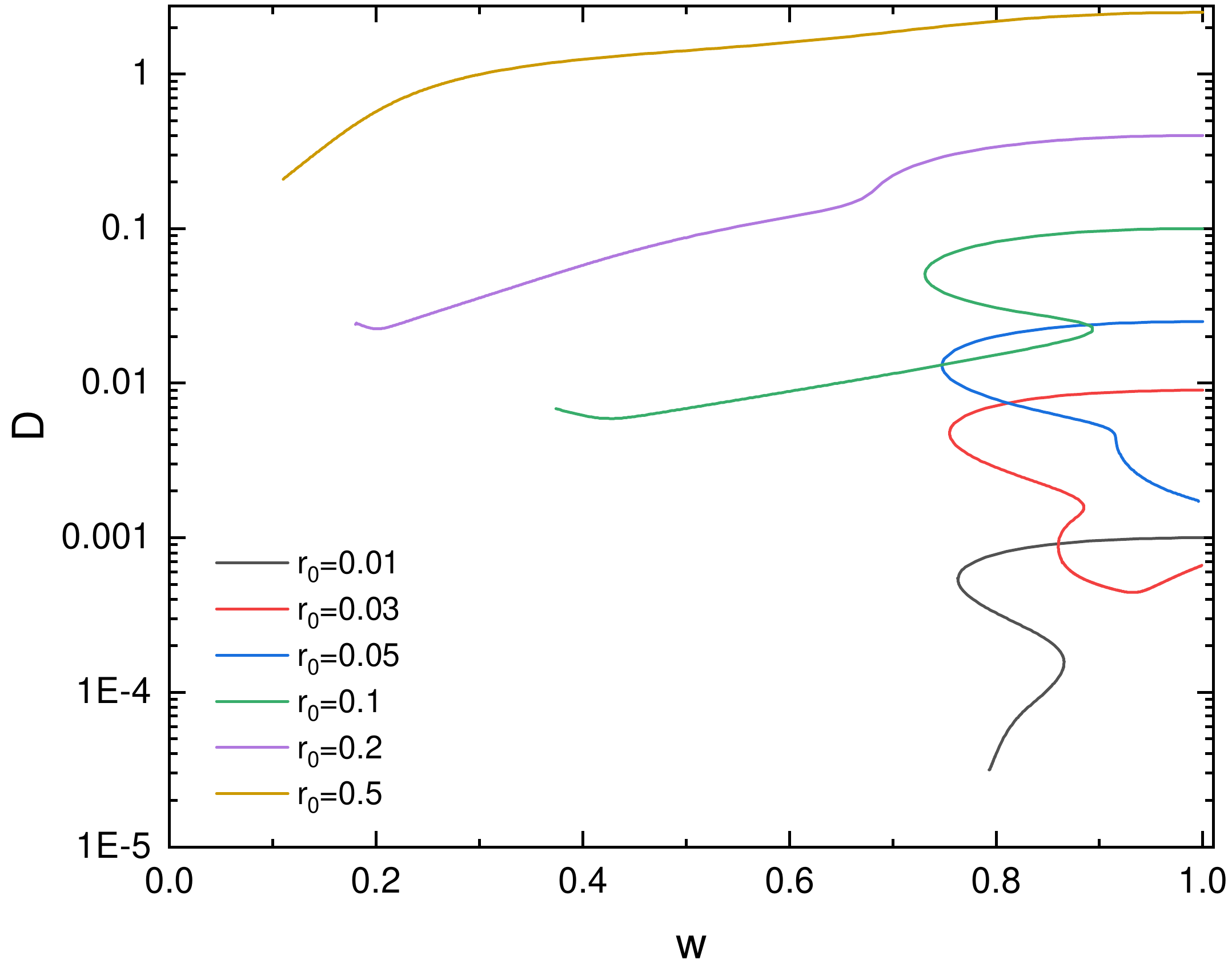}
  \end{center}
  \caption{The scalar charge $\cal D$  of  the phantom field as a function of the   frequency $\omega$ with several values of the  throat size $r_0$.
   }\label{phase3}
\end{figure}

In Fig. \ref{phase3}, we exhibit the scalar charge $\cal D$ versus
the frequency $\omega$ with the different values of  throat size $r_0$.
We can see that as $r_0$ increases, the scalar charge  also increases, which means that more phantom matter is needed.  For curves with larger $r_0$ values, the charge decreases monotonically with decreasing frequency $\omega$.

\subsection{Geometric properties of wormhole }

\begin{figure}[]
  \begin{center}
  \includegraphics[width=11cm]{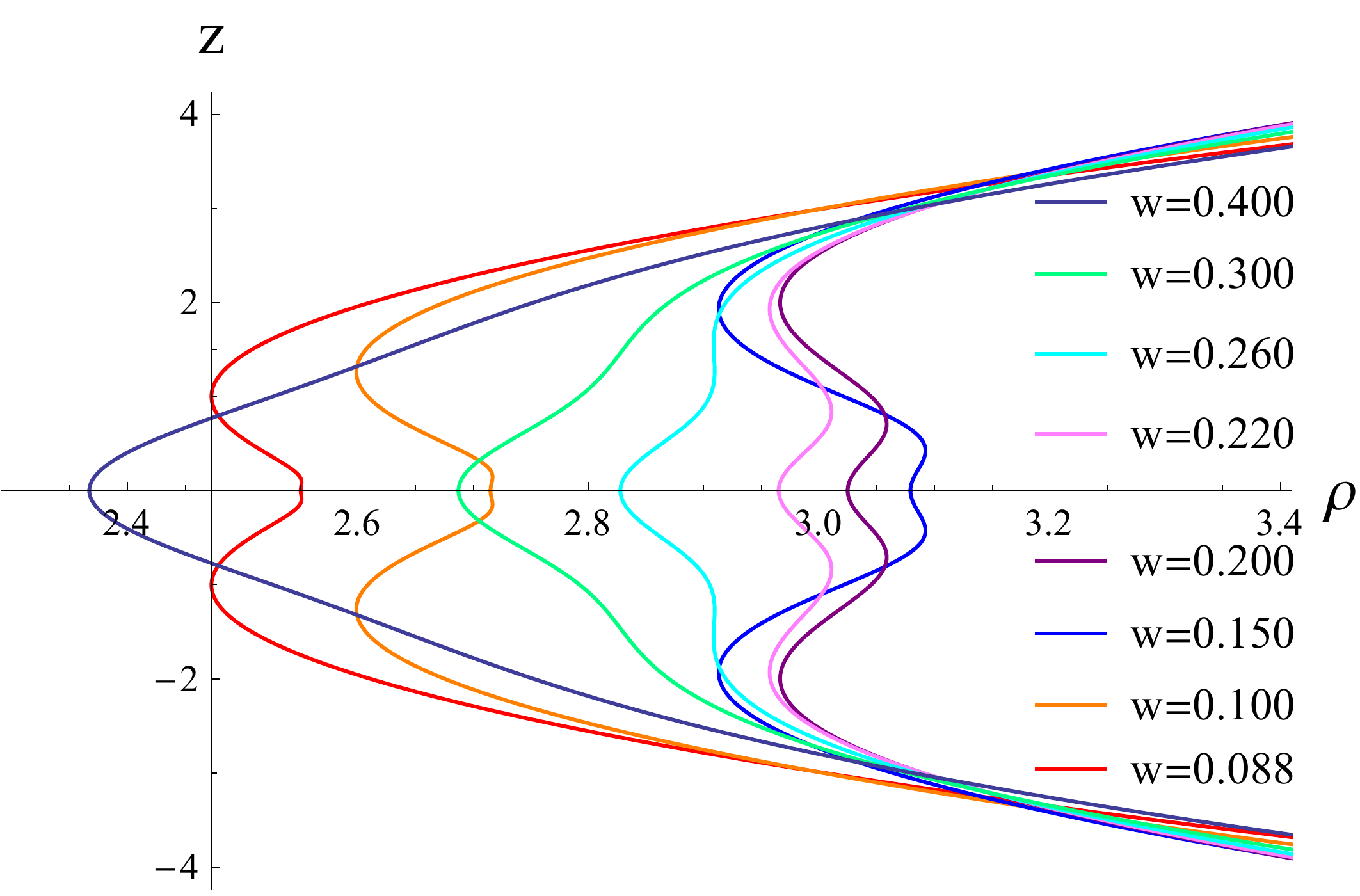}
  \end{center}
  \caption{ Geometric properties of  throats:  Two dimensional view  of  the isometric embedding of  the equatorial plane of symmetric solutions with throat parameter $r_0=1$  and several values of the complex scalar frequency $\omega$. }\label{tutr}
\end{figure}

\begin{figure}[]
  \begin{center}
  \mbox{\hspace{0.2cm}
  \subfigure[][]{\hspace{-0.0cm}
  \includegraphics[width=6cm]{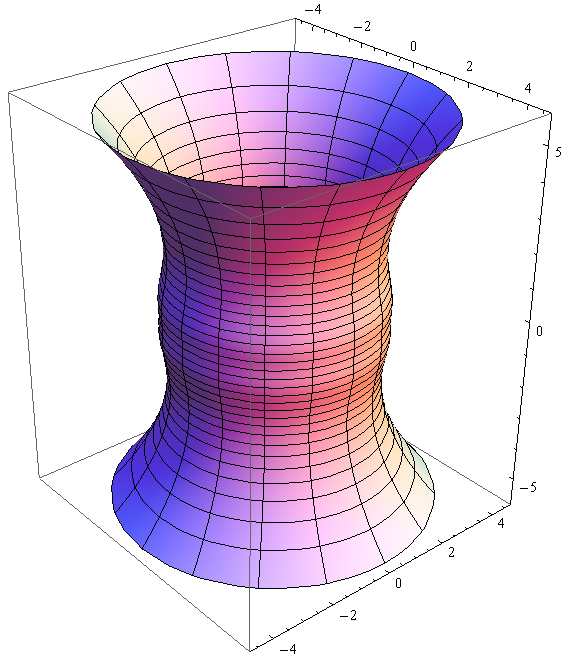}\label{Fig7c}}
  \subfigure[][]{\hspace{1.0cm}
    \includegraphics[width=6cm]{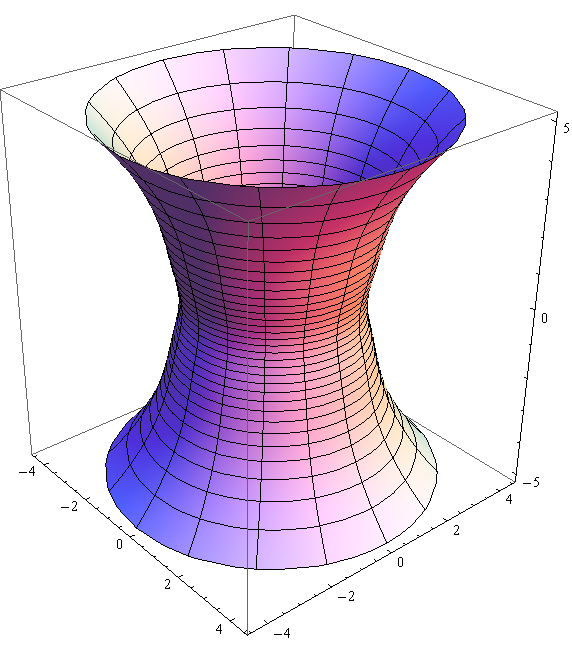}\label{Fig7d}}
    }
  \end{center}
  \caption{
  Geometric properties of  throats:  Isometric embeddings of the equatorial plane of symmetric solutions
with throat parameter $r_0=1$. (a) Three dimensional plot  with three throats
for  $\omega=0.15$. (b)  Three dimensional  plot  with  a single throat for $\omega=0.4$.}\label{phse}
\end{figure}

To explain that the metric  (\ref{line_element1})    does indeed describe a wormhole, one can make use of geometrical embedding diagrams. It is helpful to consider a two-dimensional space at a fixed time and  $\theta$. The resulting two-dimensional  spatial hypersurface of the wormhole spacetime can then be embedded in a three-dimensional Euclidean space, where the embedding diagram can be used to visualize the wormhole geometry. This technique allows us to better understand the topology and properties of the wormhole solution.

To characterize the geometry of the throat, we begin by examining the embeddings of planes with $\theta=\pi/2$.
Using cylindrical coordinates $(\rho,\varphi,z)$,
The metric on this plane can be expressed by the following formula
\begin{align}
ds^2 &= B e^{-A}  d r^2 + B e^{-A} h   d\varphi^2 \, \\
&= d \rho^2 + dz^2 + \rho^2 d \varphi^2   \,.
%&=   \left[ 1+ \left(  \frac{dz}{d \rho}   \right)^2    \right]  d \rho^2 + \rho^2 d \varphi^2  \,.
\end{align}
Comparing the two equations above,  we then obtain the expression for $\rho$  and $z$,
\begin{equation} \label{formula_embedding}
 \rho(r)= \sqrt{ B(r) e^{-A(r)} h(r) } ,\;\;\;\;\;\;\;\;\;\;   z(r) = \pm  \int  \sqrt{ B(r) e^{-A(r)}  -   \left( \frac{d \rho}{d r} \right)^2    }     d r \;.
\end{equation}
Here $\rho$  corresponds to  the circumferential radius, which corresponds to the radius of a circle located in the equatorial plane and having a constant coordinate $r$.
The function  $\rho(r)$ has extreme points, where the first derivative is zero. When the second derivative of the extreme point is greater than zero, we call this point an throat, which
corresponds to a minimal surface. When the second derivative of the extreme point is less than zero, we call this point a equator, which
corresponds to a maximal surface.

To  illustrated geometric properties of  throats, In Fig. \ref{tutr},  we show two dimensional view  of  the isometric embedding of  the equatorial plane of symmetric solutions with throat parameter $r_0=1$  and several values of the complex scalar frequency $\omega$.
When the frequency is high, the wormhole has only one throat with a small radius. As the frequency increases, the radius of the throat also increases. At near $\omega=0.26$, two new throats begin to appear, and the wormhole then has three throats.
As the frequency further increases, the radii of the three throats will first increase and then decrease. In particular, the middle throat will become less and less visible, although it always exists, but tends to disappear.
In order to get further insight into the geometry of the wormholes ,   we present two three-dimensional  plot in Fig. \ref{phse}, including one with three throats and two equators  with $\omega=0.15$, and the other with only one throat with $\omega=0.4$.
\section{Conclusions}
In this paper, we reconsidered the model of  a complex boson field and a phantom field  minimally coupled to  Einstein gravity and  investigated the domain of existence and physical characteristics of wormhole solutions embedded in  a complex bosonic matter field with  parity-odd symmetry.
We have presented the relationship between the conserved charges and frequency, which depends on the throat size. Furthermore, we have compared our results to those of the single boson star solution. Surprisingly, the curve does not exhibit a spiral shape. Instead, as the throat size $r_0$ increases from zero, the curve initially forms a loop structure. As $r_0$ continues to increase, the multi-valued curve gradually becomes a single-valued curve. Unlike the even-symmetric case studied previously, the  solutions  with  a complex odd-symmetric bosonic matter field    cannot degenerate into boson stars.

There are some interesting extensions of our work which we plan to investigate in future
projects. First, in order to achieve a wormhole solution in general relativity,  the presence of exotic matter is required.
However, under modified gravity models, it is possible to construct wormhole solutions without the need for exotic matter. Therefore, studying the existence of such solutions under modified gravity is  interesting.  Furthermore, we are planning to investigate the stability of these solutions. Finally, considering that boson stars have both ground state and excited state solutions,  we  can explore the excited state solutions in the future.

\section*{Acknowledgements}
%%%%%%%%%%%%%%%%%%%%
%%%%%%%%%%%%%%%%%%%%
YY is supported by introduction of talent research project of Northwest Minzu University (No. xbmuyjrc2020030).
YQW is supported by National Key Research and Development Program of China
(Grant No. 2020YFC2201503) and the National Natural Science Foundation of China (Grant
No. 12275110 and 12247101)

\end{document}